\documentclass[page-classic]{epl2} 
\usepackage{amsmath}
\usepackage{amssymb}
\usepackage{epsfig,graphics}
\usepackage{color}
     \definecolor{darkred}{rgb}{0.75,0,0}
     \definecolor{darkgreen}{rgb}{0,0.5,0}
     \definecolor{darkblue}{rgb}{0,0,0.75}
     \definecolor{darkorange}{rgb}{1,0.9,0.1}		
\def\hlambda{\hat{\lambda}}

\title{Stochastic Dynamics of Electrical Membrane 
with Voltage-Dependent Ion Channel Fluctuations}
\shorttitle{Electrical Membrane Stochastic Dynamics} %Insert here a short version of the title if it exceeds 70 characters

\author{Hong Qian\inst{1} \and Xue-Juan Zhang\inst{2} \and Min Qian\inst{3}}
\shortauthor{H. Qian \etal}

\institute{                    
  \inst{1} Department of Applied Mathematics, University of Washington,
Seattle, WA 98195-3925, USA;
College of Mathematics, Jilin University, Changchun, Jilin 130012, PRC\\
  \inst{2} Department of Mathematics, Zhejiang Normal University,
Jinhua, Zhejiang 321004, PRC\\
  \inst{3} School of Mathematical Sciences, Peking University,
Beijing 100871, PRC
}


\abstract{A Brownian ratchet like stochastic theory for the electrochemical
membrane system of Hodgkin-Huxley (HH) is developed.  The system 
is characterized by a continuous variable $Q_m(t)$, representing 
mobile membrane charge density, and a discrete variable $K_t$ representing ion 
channel conformational dynamics.  A Nernst-Planck-Nyquist-Johnson
type equilibrium is obtained when multiple conducting ions  have a common 
reversal potential.  Detailed balance yields a previously unknown 
relation between the channel switching rates and membrane capacitance,
bypassing Eyring-type explicit treatment of gating charge kinetics. 
From a molecular structural standpoint, membrane charge 
$Q_m$ is a more natural dynamic variable than 
potential $V_m$; our formalism treats $Q_m$-dependent 
conformational transition rates $\lambda_{ij}$ as intrinsic 
parameters.  Therefore in principle,  $\lambda_{ij}$ vs. $V_m$
is experimental protocol dependent, {\em e.g.}, 
different from voltage or charge clamping measurements.
For constant membrane capacitance per unit
area $C_m$ and neglecting membrane potential
induced by gating charges, 
$V_m=Q_m/C_m$, and HH's formalism is recovered. The
presence of two types of ions, with different channels and reversal
potentials, gives rise to a nonequilibrium steady state with 
positive entropy production $e_p$.  For rapidly fluctuating channels,
an expression for $e_p$ is obtained.
}

\begin{document}

\maketitle

\section{Introduction}

\begin{figure}[h]
\centerline{
\includegraphics[width=2.5in]{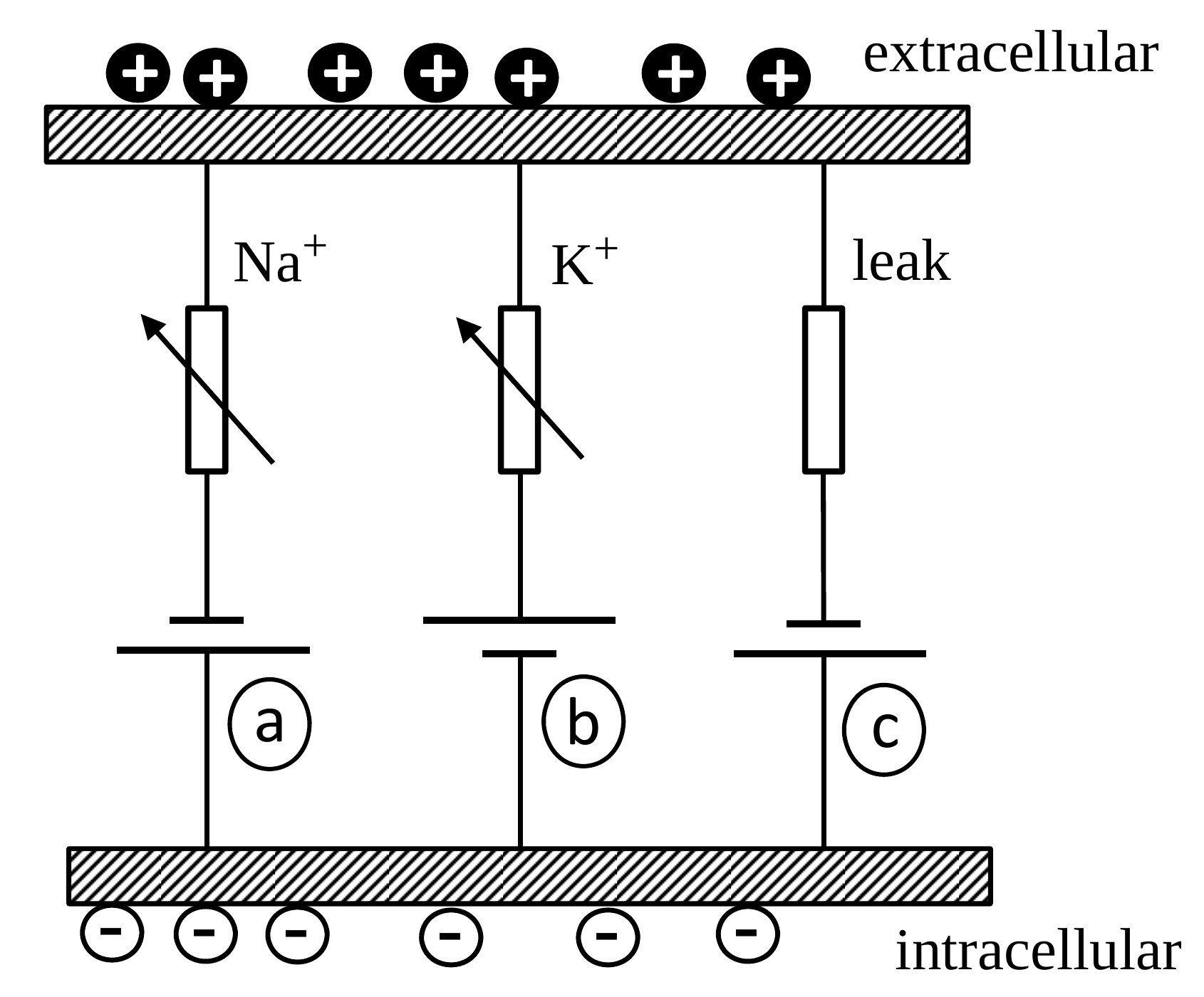}
}
\caption{A standard electrical analogue for HH's 
electrical membrane with a membrane capacitor
(denoted by the two plates) and three types of channels:
Na$^+$, K$^+$, and leak, the first two having variable
conductances.  The three batteries \textcircled{a}, 
\textcircled{b}, and \textcircled{c} 
represent the reversal potentials for the three respective
ions.  This electrical diagram shows
that if all three batteries are having a same voltage, 
then the entire system has no current, thus in an equilibirum.
However, if any two batteries are having different 
voltages, then there will be a current going through one of
the circuit loops, thus the system is in a nonequilibrium
steady state.
}
\label{fig_1}
\end{figure}

	Hodgkin-Huxley's 1952 (HH) theory of neuronal electrical 
impulse generation \cite{hh}, in terms of voltage-dependent
Na$^+$ and K$^+$ ion channels embedded in cell membrane
that is treated as a leaky capacitor, remains one of the greatest 
biophysical understandings of living processes \cite{hille_book}. 
Following the fundamental observations that 
individual ion channels stochastically fluctuate within discrete
conducting states \cite{ns_book,gating} and recent laboratory
experiments on voltage-dependent channel conformational
changes \cite{ionchannelbiol,wanglg,membrane_exp}, the HH theory has been 
extended to a mesoscopic scale \cite{lecar,defelice}.
While the literature on stochastic HH dynamics is rapidly growing
\cite{sr,shea-brown}, none has provided it a sound statistical physics
formulation including equilibrium electrochemical fluctuations as one
of its appropriate limits. 
One notices that action potentials are highly nonlinear and far from
nonequilibrium phenonmenon driven by sustained
Na$^+$ and K$^+$ ion concentration differences across excitable, 
living cell membrane \cite{deweer,hh,hille_book}.  However, when such
a driving force is absent, a passive lipid bilayer with fluctuating
ion channels should observe Nernst-Planck (NP) equilibrium and 
obey
Nyquist-Johnson's (NJ) charge fluctuations
\cite{nyquist,kittle_book}.  A complete biophysical theory should be
able to account for both equilibrium and nonequilibrium steady state (NESS)
excitable behaviors \cite{ness}.   In cell physiological terms:
There is no need for an active pump to sustain a NP-NJ 
equilibrium under a common reversal potential $V_r$.
However, there is a continuous dissipation when there are 
{\bf\em two}, different reversal potentials $V_{rK}$ and $V_{rNa}$.  
A difference between two reversal potentials
has to be sustained by an active pump such as Na$^+$/K$^+$-ATPase
\cite{deweer}.  See Fig. \ref{fig_1} for an illustration.

	This Letter contains two parts.  First, it reports a stochastic 
theory for a passive membrane,
as a capacitor with currents leaking through equilibrium
fluctuating membrane channel proteins.  To simplify 
the presentation, but without loss of generality, we consider only
one type of cation in the electrolyte aqueous solution that surround 
the membrane.  To be consistent with electrochemical NP equilibrium,
such a formulation requires a fluctuation-dissipation relation,
or detailed balance \cite{nyquist,ness,fdr_db}.  This yields a previously 
unknown relation between the  voltage-dependent 
channel conformational transition rates $\lambda_{ij}$ and membrane 
charge $Q_m$ that consists of only mobile ions, excluding
the confined charges within membrane proteins. Secondly, with the 
setup developed, a thermodynamically rigorous nonequilibrium 
formulation of stochastic HH theory with two types of  
cations is presented.  Entropy production in NESS is studied.

\section{Stochastic theory of fluctuating membrane charges}

There are electrostatic interactions between the charged groups
within proteins embedded in a lipid bilayer, channels or not, 
and the mobile electrical 
charge density on the membrane, $Q_m$:  The conformational
states of a channel protein influence the membrane charge
via a cross-membrane ionic current and intra-membrane
charge/dipole movements, and in return, the membrane 
charges affect the conformational dynamics of the proteins
through their voltage sensors, known collectively as
``gating charge'' \cite{gating,ionchannelbiol,wanglg}.  Following NJ's theory, we 
put this verbal description into a mesoscopic stochastic dynamic
model for the electrochemical system \cite{kittle_book}:
\begin{equation}
    R_{K_t}\frac{dQ_m}{dt} + \left(\frac{Q_m}{C_{K_t}}
             +\epsilon_{K_t}-V_r\right) = \sqrt{2k_BT R_{K_t}} \xi(t),
\label{eq1}
\end{equation}
in which $\xi(t)$ is a Brownian white noise, and 
$K_t$ is a discrete-state Markov jump process
collectively representing conformations of the channel 
proteins.  When $K_t=i$, the per unit area membrane 
capacitance and conductance are $C_i$ and $R_i$ respectively.
The reversal voltage $V_r$ represents an equilibrium NP potential 
for the particular conducting ions (see below).
Just as the $V_r$ is an ``effective potential'' due to the ionic 
concentration gradient across the membrane, the $\epsilon_i$ 
is an ``effective electrical field potential'' induced by
the net gating charge(s), or any other confined charge(s) in 
membrane proteins \cite{fn1}.  In the present work, the
membrane with embedded proteins is modelled as two 
conducting plates with confined charges in between. In principle,
even in the absence of $V_r$, there is a membrane potential
$\epsilon_i$;  in reality, this effect is miniscule.

The significance of this dynamic description is that it unifies 
the theories of NJ on membrane electrical fluctuations \cite{nyquist}
and HH on action potentials
\cite{hh}.  In fact, denoting membrane potential $V_m=Q_m/C$,
conductance $g=R^{-1}$, and neglecting the fluctuating
$\xi(t)$ and small $\epsilon_{K_t}$, then Eq. \ref{eq1} becomes 
$CdV_m/dt +  g(V_m)(V_m-V_r) = 0$, 
which is the starting point of HH \cite{lecar,hh}.
Several remarks for  Eq. \ref{eq1} are in order.

($i$)  It is important to point out that the $Q_m$ is {\em not}
the conducting charges across the membrane; rather
it is the membrane charge density.  See \cite{fn2} for a
derivation of (\ref{eq1}).  The physically correct way is to 
start from current  formulation as NJ and HH did, not voltage.

($ii$)  It is assumed that channel conformational 
transitions are ``instantaneous'' without being coupled to a 
$Q_m$ {\em change}: $\Pr\{K_{t+\Delta t}=j|K_t=i,Q_m(t)=q\}$ $=$ 
$\lambda_{ij}(q)\Delta t + o\big(\Delta t\big)$.
In chemomechanical systems like motor proteins, this type
of model is called {\em Brownian ratchet} \cite{ratchet}. 
It excludes the possibility of a tightly coupled simultaneous 
channel conformational transition {\it and} change in membrane 
$Q_m$, the so called {\em power stroke} scenario in molecular motor.
In mathematical physics, (\ref{eq1}) is known as a coupled (or switching) 
diffusion process \cite{coupled_diffusion}.

($iii$)  The problem of stochastic HH dynamics in terms of $\big(Q_m(t),K_t\big)$ is different from the channel conformational 
dynamics with gating charge \cite{ionchanneltheory}.    
The latter addresses a single protein dynamics in 
terms of its structure and energy landscape; 
it employs Eyring-Kramers type of formalism
\cite{gating} with which detailed balance is satisfied 
automatically.  The HH theory, however, is a
{\em phenomenological} model of an electrochemical {\em system}
that consists of both channels and a membrane immersed in an
electrolyte solution.  The transition rate
$\lambda_{ij}(q)$ represents a macromolecular conformational
change in the system's
setting.   Because of this, the 
detailed balance, or fluctuation-dissipation relation, has to be 
enforced as an additional requirement to the stochatic equation
(\ref{eq1}), as in the earlier work of Einstein and Nyquist \cite{fdr_db},
and studies on motor proteins \cite{qian_jpc_02}.  

($iv$) Indeed, a quantitative isomorphism exists between an 
electrochemical system and a mechanochemical system
such as molecular motors and ligand-receptor 
dissociation kinetics under an external 
force \cite{qian_ps_02}.

($v$)  It is assumed that gating charge movements within a
channel affect the membrane capacitance as well as changing
its permeability for the conducting ions.   Therefore, with 
the $Q_m$ remains constant immediately before and after 
the conformational transition, membrane potential jumps
from $(Q_m/C_i+\epsilon_i)$ to $(Q_m/C_j+\epsilon_j)$.  
This  is the response to gating sensor movements associated 
with the conformational transition.  The present theory, therefore,
provides a quantitative link between molecular physics model
of channel gating and the  HH theory.  Its validity 
can and should be experimentally tested.

($vi$) The $V_r$ accounts for a concentration 
difference across the membrane that is in equilibrium 
with a membrane potential.  According to NP
equation, an ionic concentration gradient across the  membrane, say
the K$^+$ with $c_1$ and $c_2$, gives rise to an equilibrium
membrane potential $V_r = \big(k_BT/e\big) \ln\left(c_1/c_2\right)$. 
In such an equilibrium, the net K$^+$ current is zero, with
the membrane potential $V_r$ driven ionic current exactly balances
the concetration gradient driven ionic flux in the opposite direction.
However, such an NP equilibrium is conditioned on
the absolutely zero counter-ion, e.g., Cl$^-$, passing through the 
membrane.  In reality, if the membrane voltage is not clamped, a slow 
counter-ion leakage gradually causes both membrane potential 
{\em and} the ion concentration difference to dissipate.  
The $V_r$ in (\ref{eq1}) can be treated as a constant only on the 
time scale of no significant counter-ion leakage.

%\revision{
($vii$) There are several possible lines of experimental
and computational tests for this theory. One is an 
integrative analysis of empirically determined voltage-dependent 
gating kinetics with computational molecular dynamics in an
ionic solution with proper handling of ions and charges.
Even though nano-scale Coulomb meter for measuring membrane
charge is still in its infancy \cite{wanglg}, being able to
probe membrane voltage and charges as separated 
physical quantities will provide a deeper understanding of 
the complex ``capacitance''.  Finally, beyond single-channel 
recording, experiments on artificial membrane with two types of 
ion channels in the presense of two different ionic reversal 
potentials will open a new vista for nonequilibrium physical biology.
%}

\section{A thermodynamic linkage relation between channel and membrane}

\begin{figure}[h]
%%%%%%%%%%%%
\begin{picture}(0,90)(-140,-10)
\put(-10,-10){$(j,q)$}
\put(70,-10){$(j,q+dq)$}
\put(-10,50){$(i,q)$}
\put(70,50){$(i,q+dq)$}
%transitions
\put(5,1){\vector(0,1){42}}
\put(96,1){\vector(0,1){42}}
\put(1,43){\vector(0,-1){42}}
\put(92,43){\vector(0,-1){42}}
\put(20,56){\vector(1,0){45}}
\put(65,52){\vector(-1,0){45}}
\put(20,-3){\vector(1,0){45}}
\put(65,-7){\vector(-1,0){45}}
%rate constants
\put(23,-30){$e^{\frac{U_j(q)dq}{k_BT}}$}
\put(100,17){$\frac{\lambda_{ji}(q+dq)}{\lambda_{ij}(q+dq)}$}
\put(25,63){$e^{-\frac{U_i(q)dq}{k_BT}}$}
\put(-32,17){$\frac{\lambda_{ij}(q)}{\lambda_{ji}(q)}$}
\end{picture}
\vskip 0.5cm
\caption{In the electrochemical theory of a membrane with
fluctuating ion channels, detailed
balance condition for equilibrium is expressed as a relation between the
membrane electrical potential functions $U_k(q)=q^2/(2C_k)$ and transition 
rates $\lambda_{ij}(q)$, as given in Eq. \ref{eq3}, over a kinetic 
cycle illustrated here.
}
\label{fig_2}
\end{figure}
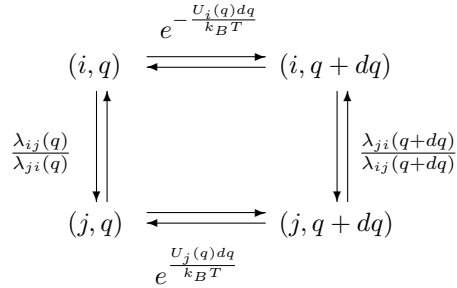

From (\ref{eq1}), 
the probability of $\big(Q_m(t),K_t\big)$, $f_{Q_mK}(q,i,t)$,
satisfies the Fokker-Planck-master equation
\begin{eqnarray}
   \frac{\partial f_{Q_mK}(q,i,t)}{\partial t} &=& \frac{\partial}{\partial q}
          \left(\frac{1}{R_i}\right)
                    \left[k_BT\frac{\partial f_{Q_mK}(q,i,t)}{\partial q} 
     + \left(\frac{q}{C_i}+\epsilon_i-V_r\right)f_{Q_mK}(q,i,t)\right]
\nonumber\\
        &&          + \sum_j \Big[ f_{Q_mK}(q,j,t)\lambda_{ji}(q) - f_{Q_mK}(q,i,t)\lambda_{ij}(q)  \Big].
\end{eqnarray}
To have an electrochemical equilibrium one has to have both
total electrical current and chemical (Fickian) flux for each species 
being zero simultaneously.  Therefore,
\begin{equation}
        \frac{\lambda_{ij}(q)}{\lambda_{ji}(q)} =
            \frac{f_{Q_mK}^{eq}(q,j)}{f_{Q_mK}^{eq}(q,i)} =
             \frac{\lambda_{ij}(0)}{\lambda_{ji}(0)}
                      \exp\left\{-\frac{1}{k_BT}\left(
                   \frac{q^2}{2C_j}+\epsilon_j q-
                   \frac{q^2}{2C_i}-\epsilon_iq\right)\right\}.
\label{eq3}
\end{equation}
We note that the  $V_r$ disappeared from (\ref{eq3}), as it
should be.   Eq. \ref{eq3} is the detailed balance  relation
for the mixed discrete and continuous  
variables  \cite{wang-oster,qian_jpc_02}, 
as illustrated in Fig. \ref{fig_2}.

\section{Voltage-dependent channel conformational transition rates}

Usually, the changes in capacitance due to gating charge moments
are much smaller than the contribution from the lipid bilayer.    
Therefore, we can assume  $C_i=C_m+\delta_i$ with
$\delta_i\ll C_m$. In this case, the term on the rhs 
of (\ref{eq3}) becomes 
$\exp\big\{-\big[q(\epsilon_i-\epsilon_j)+q^2(\delta_i-\delta_j )/(2C_m^2)\big]/(k_BT)\big\}$.  Therefore, in 
terms of the membrane potential $V_m=q/C_m$ one has
\begin{equation}
		 \frac{\lambda_{ij}(V_m)}{\lambda_{ji}(V_m)} =
               \frac{\lambda_{ij}(0)}{\lambda_{ji}(0)}
                      \exp\left\{-\frac{1}{k_BT}
                       \left( (\epsilon_i-\epsilon_j)C_mV_m 
                + \frac{(\delta_i-\delta_j)V_m^2}{2}\right) \right\}.
\label{eq5}
\end{equation}
The $(\epsilon_i-\epsilon_j)C_m$ term and  the $(\delta_i-\delta_j)$ 
term  are due to charge movements and changes in dipole moments 
associated with gating, the conformational change.  See Hill and Chen 
\cite{ionchanneltheory} for more discussions.  If $\delta_i$ is
not much smaller than $C_m$, then $V_m$ is discontinuous
and  the ``voltage-dependent rate'' has to be defined as 
$\hlambda_{ij}(V_m)\equiv \lambda_{ij}(C_iV_m)$.

	The membrane capacitance $C_m$ appearing on the
rhs of (\ref{eq5}) should not be a surprise.  ``voltage-dependent
transition rates'' actually depends upon the nature of 
applied force law, e.g., elastic cantilever or constant
force clamping \cite{cole_qian}.  This means
the voltage-dependent transition rates in HH theory, in 
principle, are different from those obtained from voltage-clamped
single-channel recording.  However, the $\lambda_{ij}(q)$ are
intrinsic properties of a membrane-channel system. 

{\bf\em  Equilibrium electrochemical fluctuations |}
With the condition given in (\ref{eq3}), one has the 
Boltzmann's law which dictates that in an thermo-electrico-chemical
equilibrium, with the stationary probability
\begin{eqnarray}
        f^{eq}_{Q_mK}(q,i) 
	&=& \mathcal{N}_1^{-1}
               \exp\left(-\frac{(q-q_i^*)^2}{2C_ik_BT
               }-\frac{\mu_i(0)-C_i(V_r-\epsilon_i)^2/2}{k_BT}\right)
\label{eqdis1}\\
	&=&  \mathcal{N}_2^{-1}
                        \exp\left(-\frac{\mu_i(q)-\mu_1(q)}{k_BT}
                         -\frac{(q-C_1V_r)^2}{2C_1k_BT}\right),
\label{eqdis2} 
\end{eqnarray}
in which $q_i^*=(V_r-\epsilon_i)C_i$, and 
\begin{equation}
      \mu_i(q) -\mu_j(q)  = k_BT \ln
                     \left(\frac{\lambda_{ij}(q)}{\lambda_{ji}(q)}\right),
\label{eqdis3}
\end{equation}
which is independent of $V_r$, and
\begin{equation}
   \mathcal{N}_1 =  \sum_i \sqrt{2\pi C_ik_BT}\ 
                         e^{-\frac{\mu_i(0)}{k_BT}
                         +\frac{C_i(V_r-\epsilon_i)^2}{2k_BT}},
\end{equation}
\begin{equation}
   \mathcal{N}_2 = \mathcal{N}_1\ e^{\frac{\mu_1(0)}{k_BT}
                               -\frac{C_1V_r^2}{2k_BT}}.
\end{equation}
The detailed balance condition (\ref{eq3}) guarantees zero
entropy production in the stationary stochastic dynamics \cite{ness}.

\section{Nonequilibrium steady state with non-zero entropy production}

For system having two types of cations with different reversal potential 
$V_{rK}$ and $V_{rN}$, and corresponding Markov channels kinetics 
$K_t$ and $N_t$, the membrane dynamics can be represented by an
equation parallel to (\ref{eq1}).  Following HH's theory, we assume
that $C_i\approx C_m$, neglecting $\epsilon$, and 
denote $V_m=Q_m/C_m$.  Then the stochastic HH equation:
\begin{equation}
    C_m\frac{dV_m}{dt} + g_{K_t}\big(V_m
             -V_{rK}\big) +
               \gamma_{N_t}\big(V_m-V_{rN} \big) 
                = \sqrt{2k_BT\big(g_{K_t}+\gamma_{N_t}\big)} \xi(t).
\label{eq10}
\end{equation}
We note that if $V_{rK}=V_{rN}$, then the two terms on the
lhs can be merged into a single term, and the results from the 
previous section apply.  However, if they are not equal,
then the stationary process is a NESS.
To prove that, we have electrical equilibrium
\begin{equation}
     f_{V_mKN}^{ness}(v,i,j) = \widetilde{\mathcal{N}}_{ij}
                        \exp\left(-\frac{C_m(v-v_{ij}^*)^2-C_mv_{ij}^{*2}}{2k_BT}\right),
\label{11}
\end{equation}
in which
\begin{equation}
         v_{ij}^* = \frac{g_{i}V_{rK}+\gamma_{j}V_{rN}}{g_{i}+\gamma_{j}},
\end{equation}
which is related to the Goldman-Hodgkin-Katz diffusion 
potential \cite{hille_book}.  Using (\ref{11}) and if the steady
state were detailed balanced between $(i,j)$ and $(i',j')$ at $v=0$,
then one would have
\begin{equation}
           \frac{f_{V_mKN}^{ness}(v,i,j)}{ f_{V_mKN}^{ness}(v,i',j') }
   = \frac{\lambda_{i'j'\rightarrow ij}(0)}{\lambda_{ij\rightarrow i'j'}(0)}
                \exp\left[ 
                         -\frac{C_m\big(v_{ij}^*-v_{i'j'}^*\big)v}{k_BT}\right].
\label{eq14}
\end{equation}
However when $V_{rK}\neq V_{rN}$, $v_{ij}^*$ is a function
of $i$ and $j$, and the rhs of (\ref{eq14}) can not be equal to 
$\lambda_{i'j'\rightarrow ij}(v)/\lambda_{ij\rightarrow i'j'}(v)$ which 
has to be independent of $V_{rN}$ and $V_{rK}$. 
This proves that the stationary process is now a NESS.  Note that
$\lambda_{ij\rightarrow i'j'}(v)$ satisfies detailed balanced for each 
given $v$ \cite{ns_book}.   The same model, therefore, can account
for both equilibrium NP-NJ theory, as well as HH's excitable 
dynamics in a NESS. 

	One can in fact obtain the NESS distribution analytically
in the case of rapid channel fluctuations \cite{qian_jpc_02}:
\begin{equation}
  f^{ness}_{V_mKN}(v,i,j) =  
      \pi(i,j|v)      \mathcal{N}^{-1}  
                   \exp\left[ -\frac{C_m}{k_BT}
                  \left( \frac{v^2}{2}- \int_0^v \varphi^*(y) dy\right) \right],
\end{equation}
in which  $\mathcal{N}$ is a normalization factor, and 
\begin{equation}
          \pi(i,j|v)\lambda_{ij\rightarrow i'j'}(v) =
             \pi(i',j'|v)\lambda_{i'j'\rightarrow ij}(v), 
\end{equation}
\begin{equation}
     \varphi^*(v) =   \frac{\sum_{i,j} \big(g_iV_{rK}+\gamma_jV_{rN}\big)
                                   \pi(i,j|v) } 
                  {\sum_{i,j} (g_i+\gamma_j)\pi(i,j|v) }.
\end{equation}
The corresponding NESS entropy production is
\begin{equation}
    e_p = \sum_{i,j}\int_{-\infty}^{\infty}
             \frac{(g_i+\gamma_j) \big(\varphi^*(v)-v_{ij}^*\big)^2}{k_BT}
               \  f_{V_mKN}^{ness}(v,i,j)dv.
\end{equation}
When $V_{rK}=V_{rN}$, $\varphi^*(v)=v_{ij}^*$ and
$e_p=0$.

\section{Summary} 

While it is widely accepted that HH's
action potential dynamics of an excitable electrical membrane
has to be a nonequilibrium phenomenon \cite{hille_book},  which 
is driven by the difference in Na$^+$ and
K$^+$ reversal potentials,  there has never been a quantitative
statistical thermodynamic theory that adequately account for
this intuition.  Using the stochastic mathematics of coupled 
diffusion widely employed in Brownian ratchet theory for molecular
motors, we have developed a quantitative model which takes in account
both NP-NJ type of charge-chemical equilibrium with fluctuations, and
stochastic HH's theory.  For the equilibrium part, we discovered 
a previously unknown relation akin to a detailed balance condition. 
We show that if there are two ions with unequal reversal potentials, 
then a NESS arises.  In the case of rapid channel fluctuation, we
obtained an explicit expression for the NESS entropy production, 
which becomes zero when the two reversal potentials are equal. 
This model provides the basis for future investigations of 
the emergence of organized living phenomenon through 
nonequilibrium physics.

\acknowledgments
We thank F. Bezanilla, I. Goychuk, 
B. Hille, Bo Li, Wei-Shi Liu, R. Nossal, D. Sigg, P.J. Thomas, 
Li-Guo Wang, T.B. Woolf and Fu-Xi Zhang for helpful discussions.  
H.Q. also thanks Profs. Paul De Weer and Luis Reuss for introducing him to membrane biophysics.


\begin{thebibliography}{0}
\bibitem{fn1}
In the general discussion on ``voltage sensing'' of 
voltage-dependent ion channels, the translocation of 
a fixed charge, or charges, and movements of dipole
moments are two mechanisms.  The term $\epsilon_{K_t}$
represents, phenomenologically, the former and 
the $C_{K_t}$ represents the latter.  We we shall show
below, they yield $V_m$ and $V_m^2$ dependence,
respectively.  See also \cite{ionchanneltheory}.
\end{thebibliography}
\end{document}